# Evaluation of centroiding algorithms for an autonomous star tracker

Márcio Afonso Arimura Fialho[1*]

[1]DIEEC, INPE, Av. dos Astronautas, 1758, São José dos Campos, SP, Brazil

[*]e-mail: marcio.fialho@inpe.br

**Abstract**

This work presents numerical results of a computer simulation performed with six centroiding algorithms targeting a star tracker in development at INPE, including readout noise and considering a Gaussian point spread function. Five of the tested algorithms are light-weight centroiding algorithms with low computational costs. These were compared to a shape fitting algorithm based on the lsqnonlin function available in Matlab and GNU Octave. The algorithms studied here are also applicable for astrometry and adaptive optics.

**Keywords**: star trackers, centroiding.

**Notice**: This is an extended and revised version of a work previously presented as a poster (CBDO 087-A) in CBDO-2024. This is also a preprint, before peer-review, of a paper submitted to EPJ-ST for possible publication.

## 1 Introduction

A star tracker is one of the most accurate attitude sensors used aboard spacecraft. It uses observed stars as reference for attitude (spatial orientation). The accuracy of a star tracker depends critically on the accuracy of the centroiding algorithm used to process images acquired by the star tracker.

INPE is developing an autonomous star tracker (AST-INPE) [1], see Figure 1. An autonomous star tracker is a star sensor capable of acquiring an attitude solution even when no a priori attitude estimate exists ("lost in space" case).

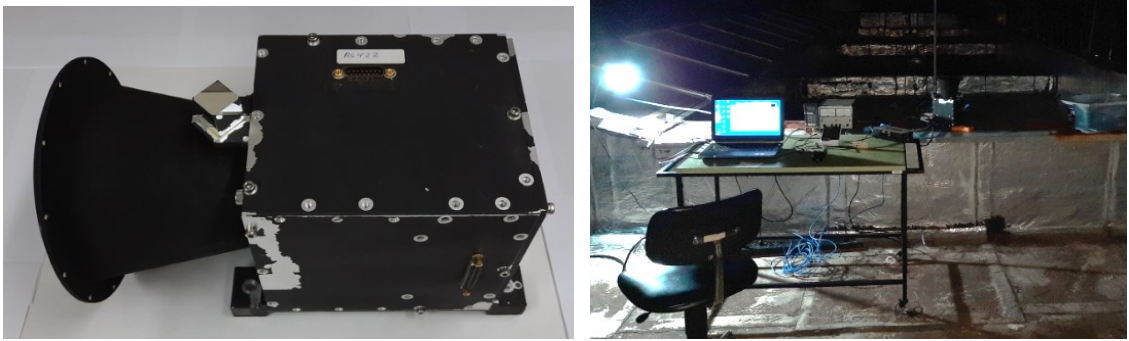

Figure 1 - left: one of the engineering models of the AST-INPE; right: night sky test with AST-INPE.

A centroiding algorithm associated with a segmentation algorithm was implemented in this star tracker. However, it was noted that this algorithm underestimates the brightness of dim stars. This motivated an evaluation of other centroiding algorithms for this star tracker.

## 2 Centroiding Algorithms

In total, six centroiding algorithms were tested, labeled ALG-1 to ALG-6 in this work.

### 2.1 ALG-1

ALG-1 is based on the simple Center of Gravity (CoG) algorithm, after subtracting the background level $b$ and considering only pixels above a threshold $T$. It is similar to the modified moment algorithm described in [2] and it is very similar to the algorithm currently implemented in AST-INPE. The centroid $(x_c, y_c)$ is computed using the following equations:

$$x_c = \frac{1}{A} \sum_{x_{min}}^{x_{max}} \sum_{y_{min}}^{y_{max}} x \cdot I'(x,y), \qquad y_c = \frac{1}{A} \sum_{x_{min}}^{x_{max}} \sum_{y_{min}}^{y_{max}} y \cdot I'(x,y),$$

with

$$A = \sum_{x_{min}}^{x_{max}} \sum_{y_{min}}^{y_{max}} I'(x,y), \qquad I'(x,y) = \begin{cases} I(x,y) - b & , I(x,y) > T \\ 0 & , I(x,y) \leq T \end{cases}$$

$$T = b + 3.5\sigma$$

where $(x, y)$ are the coordinates of the pixel center, $I(x,y)$ is the raw pixel intensity, $I'(x,y)$ is the corrected pixel intensity, $A$ is the estimated star brightness in digital levels, $b$ is the estimated background level and $\sigma$ is the standard deviation of the image background level.

### 2.2 ALG-2

ALG-2 is similar to ALG-1, the difference being that all pixels inside the centroiding window are considered, so that no thresholding is performed (this is equivalent to set $T < 0$).

### 2.3 ALG-3

ALG-3 is based on the Iteratively Weighted Center of Gravity (IWCoG) algorithm [3]. At each iteration $i$, the corrected pixel intensity $I'(x,y)$ used in the equations that compute $x_c$, $y_c$ and $A$ in ALG-2, is replaced by $I''(x,y) = W_i(x,y) \cdot I'(x,y)$, where $W_i(x,y)$ is a weighting function computed at each iteration, following a normal distribution:

$$W_i(x,y) = \exp\left[-\left\{\frac{(x - x_{c,i-1})}{2\sigma_{sh}^2} + \frac{(y - y_{c,i-1})}{2\sigma_{sh}^2}\right\}\right],$$

with $x$ and $y$ being the coordinates of the pixel center, $(x_{c,i-1}, y_{c,i-1})$ the centroid computed in the previous iteration and $\sigma_{sh}$ a parameter that defines the "width" of the weighting function, chosen in the Monte Carlo simulations (below) as 0.7 pixel. In the first iteration, the initial centroid $(x_{c,0}, y_{c,0})$ that is used to compute $W_1(x,y)$ is taken as the geometric center of the centroiding window. In the simulations, centroiding was iterated 10 times.

## 2.4 ALG-4

ALG-4 is based on the Intensity Weighted Centroiding (IWC) [3, 4]. It is very similar to ALG-2, but with the corrected pixel intensity $I'(x, y)$ used in the equations that compute the centroid replaced by $I'(x, y)$ raised to the power $q$, where $q$ is a small positive number larger than 1, usually chosen around 2. In the simulations $q$ was chosen as 2.0.

## 2.5 ALG-5

ALG-5 combines the exponent $q$ of the IWC with the thresholding of ALG-1, that is, it is a version of ALG-4 that performs thresholding like ALG-1 instead of using all the pixels as in ALG-2. In the Monte Carlo simulations $q$ was chosen as 2.0.

## 2.6 ALG-6

ALG-6 is a shape fitting algorithm based on the least squares method that fits a Gaussian shape to the observed (or simulated in case of Monte Carlo simulations) star image. It is based on the Matlab `lsqnonlin` function from the Optimization Toolbox. In the simulations, for each pixel the Gaussian shape was sampled in a 4x4 uniformly spaced sub-grid and then summed.

# 3 Methodology

The algorithms were evaluated through Monte Carlo simulations. In these simulations, synthetic images measuring 7x7 pixels were created, with a single star near its center, using the parameters derived from AST-INPE. The star's Point Spread Function (PSF) was modeled with a Gaussian distribution. Even though the PSF is never truly Gaussian, this distribution provides a fairly good approximation for well focused images near the optical axis. Figure 2 presents the test program and the star image from where the parameters used in the simulations were derived: ($\sigma_{PSF} = 0.663\ px$; $A_{ref} = 10{,}046.34$; $bkg\_level = 56.84$; $\sigma_{readout\ noise} = 108\ e^-$; $1\ DN = 100\ e^-$). To get more accurate results, for each pixel the Gaussian PSF was sampled in a 10x10 uniformly spaced sub-grid and the results summed.

For each of these synthetic images, the 24 pixels at the edges of the image were used to estimate the background level $b$ and the threshold $T$ when needed. The internal 5x5 window was used as the centroiding window for the selected centroiding algorithm.

Shot-noise, read-out noise, A/D conversion truncation error and A/D saturation were modeled. Column non-uniformity and hot/warm pixels present in images generated by AST-INPE were not modeled.

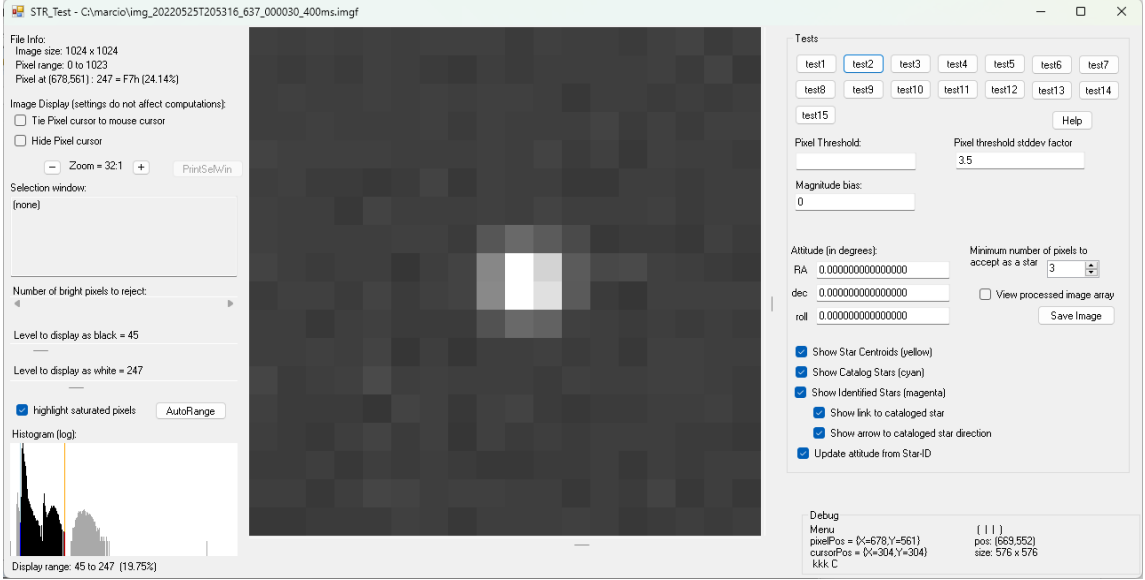

**Figure 2 - Test program used to analyze images acquired by AST-INPE, showing the star image ($m_V = 2.65$) used to derive parameters for the Monte Carlo simulations.**

For each synthetic image, the true centroid position $(x_t, y_t)$ was taken at random in the central pixel, with coordinates uniformly distributed in this pixel, that is: $x_t \in (3,4)$ and $y_t \in (3,4)$.

The estimated brightness $A$ computed by the centroiding algorithms was used to compute the magnitude, using the following expression:

$$mag_{est} = -2.5 * \log_{10}\left(\frac{A}{A_{ref}}\right),$$

with $A_{ref}$ being the digital level of a star with zero magnitude.

The computed centroids and estimated magnitudes were compared to the true centroid locations and true magnitudes (known in the simulation) to compute the centroiding and magnitude estimation errors:

$$err_{ctr} = \|(x_c, y_c) - (x_t, y_t)\|, \qquad mag_{err} = mag_{est} - mag_{true}$$

## 4  Results

Figures 3 and 4 present the centroiding error and magnitude errors, for 10,000 runs for each datapoint. Magnitude errors were not calculated for ALG-3, ALG-4 and ALG-5, but they should be the same as those of ALG-2, ALG-2 and ALG-1, respectively.

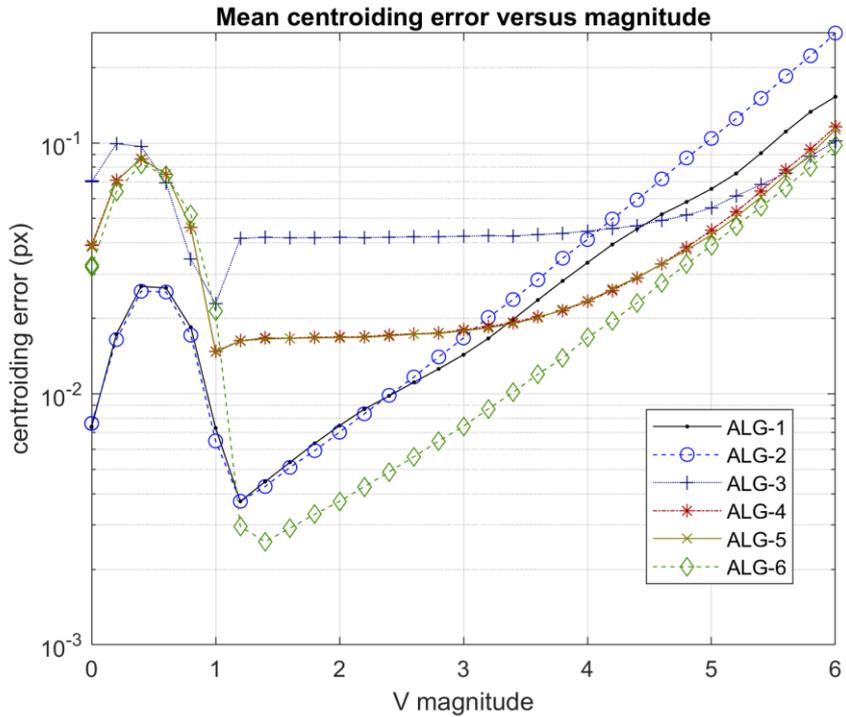

Figure 3 – Centroiding errors (in pixels) versus magnitudes in Johnson's V band.

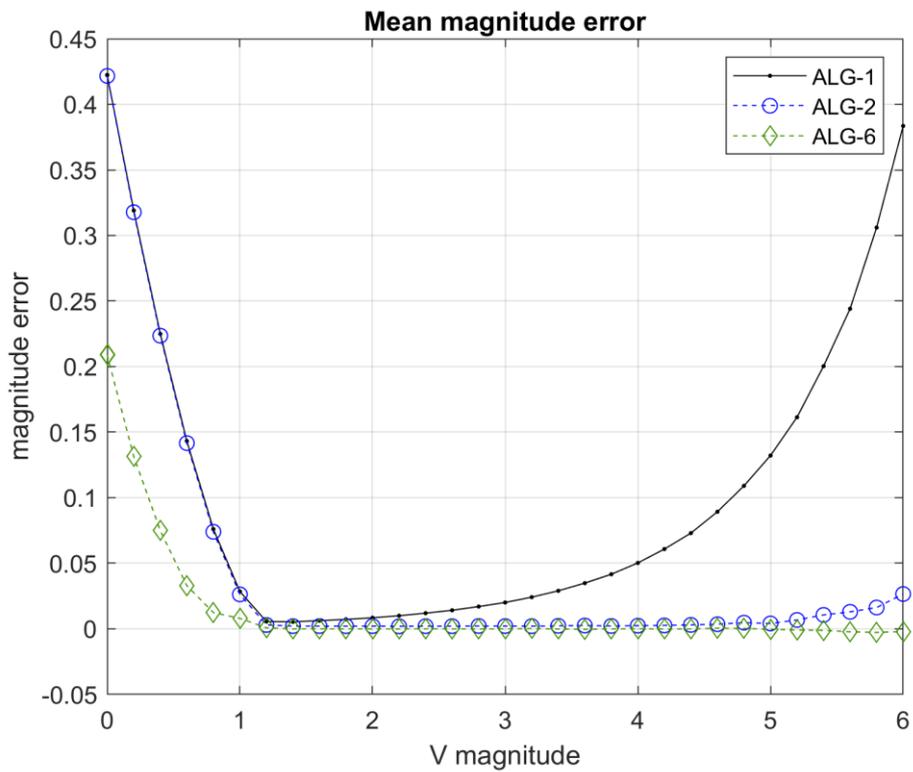

Figure 4 – Magnitude errors versus magnitudes in Johnson's V band.

Table 1 presents execution time measurements for each algorithm, considering stars of magnitude 5.0. These measurements also include the time spent in generating the simulated image. In the first column, a dummy algorithm that immediately returns the true centroid was included, so that the time spent in generating synthetic images could be appreciated. Tests were performed with Matlab 2024b running in a modern notebook with CPU at 3.6 GHz.

Table 1 - Average time spent for each run in microseconds, including time used to generate a 7x7 pixel synthetic image with a magnitude 5 star.

| Algorithm | dummy | ALG-1 | ALG-2 | ALG-3 | ALG-4 | ALG-5 | ALG-6 |
|---|---|---|---|---|---|---|---|
| average time* (μs) | 148.9 | 154.5 | 156.2 | 163.5 | 166.5 | 153.1 | 5609.2 |
| avg. time std. dev.* (μs) | 10.9 | 11.7 | 12.9 | 17.9 | 31.3 | 8.8 | 775.0 |

\* Computed by dividing the 10,000 runs into 5 lots of 2000 runs each, measuring the time spent to run each lot, computing the mean and standard deviation from these 5 measured times and dividing them by 2000.

## 5 Summary of results and discussion

In Figures 3 and 4 we can see a large increase in centroiding and magnitude estimation errors for stars brighter than magnitude 1.2. This is due to saturation of the central pixels at a DN of $2^{10} - 1 = 1023$ for very bright stars. For intermediate magnitudes (1.4 to 4.0) ALG-3, that is based on the IWCoG algorithm, performs poorly. This was also reported by [4]. ALG-4 and ALG-5 also behave similarly to ALG-3 for intermediate magnitudes. For dim stars (mag > 4), the most common stars in star tracker images, ALG-2 performs the worst, followed by ALG-1. The large increase in magnitude errors for dim stars noticed for ALG-1 is due to thresholding, as for these dim stars a large fraction of the signal is in pixels below the threshold $T$. In AST-INPE this thresholding is a side effect of the segmentation algorithm used to separate star images from the background. ALG-6, a shape fitting algorithm, is the most accurate, both in centroiding performance and in magnitude estimation, but it has the largest computational cost of all algorithms tested. For ALG-1 to ALG-5 the computational cost could not be properly evaluated, since for these algorithms the computational cost was dominated by the generation of synthetic images.

### 5.1 Limitations and future work

It should be noticed that, for simplicity, this work considers a Gaussian PSF in the simulations. However in practice the PSF can depart significantly from a Gaussian distribution, meaning that the results obtained in this work in terms of centroiding accuracy are not necessarily applicable in every practical situation. For example, in a given scenario the best performing centroiding algorithm in terms of accuracy could be different. This work could be extended by running the simulations again, considering a more realistic PSF.

## 6 Acknowledgments

We thank FINEP, CNPq, FUNCATE and FUNDEP for the financial support, and for all collaborators who have been involved in this project. Without their help this project would not materialize.

# 7 Code availability

The code developed by the author for this work will be made public in due couse. Readers interested in obtanining a copy of the code before it is made public can make a request by sending an e-mail to the author.